\documentclass[preprint,preprintnumbers,amsmath,amssymb]{revtex4}
\usepackage{amsfonts}    
\usepackage{amssymb}
\usepackage{latexsym}
\usepackage{graphicx}
\usepackage{verbatim}
\usepackage{rotating}
\usepackage{multirow}
\usepackage[usenames,dvipsnames]{color}
\usepackage[active]{srcltx}

\newcommand{\al}{\alpha}

\newcommand{\la}{\lambda}

\newcommand{\Om}{\Omega}

\newcommand{\de}{\delta}

\newcommand{\De}{\Delta}

\newcommand{\tha}{\theta}

\newcommand{\rar}{\rightarrow}
\newcommand{\lrar}{\leftrightarrow}

\begin{document}

\title{Few-electron atomic ions in non-relativistic QED: the Ground state}

\author{Alexander~V.~Turbiner}
\email{turbiner@nucleares.unam.mx, alexander.turbiner@stonybrook.edu}
\author{Juan Carlos Lopez Vieyra}

\email{vieyra@nucleares.unam.mx}
\affiliation{Instituto de Ciencias Nucleares, Universidad Nacional
Aut\'onoma de M\'exico, Apartado Postal 70-543, 04510 M\'exico,
D.F., Mexico{}}

\author{Horacio~Olivares-Pil\'on}
\email{horop@xanum.uam.mx}
\affiliation{Departamento de F\'isica, Universidad Aut\'onoma Metropolitana-Iztapalapa,
Apartado Postal 55-534, 09340 M\'exico, D.F., Mexico}

\begin{abstract}
Following detailed analysis of relativistic, QED and mass corrections for
helium-like and lithium-like ions with static nuclei for $Z \leq 20$
the domain of applicability of Non-Relativistic QED (NRQED) is localized
for ground state energy. It is demonstrated that for both helium-like and
lithium-like ions with $Z \leq 20$  the finite nuclear mass effects do not change
4-5 significant digits (s.d.), and the leading relativistic and QED effects leave
unchanged 3-4 s.d. in the ground state energy. It is shown that the
non-relativistic ground state energy can be interpolated with accuracy not
less than 13 s.d. for $Z \leq 12$, and not less than 12 s.d. for $Z \leq 50$ for helium-like
as well as for $Z \leq 20$ for lithium-like ions by a compact meromorphic function in variable
${\lambda}=\sqrt{Z-{Z_B}}$ (here $Z_B$ is the 2nd critical charge \cite{TLO:2016}),
$P_9(\lambda)/Q_5(\lambda)$.
It is found that both the Majorana formula - a second degree polynomial
in $Z$ with two free parameters - and a fourth degree polynomial in ${\lambda}$
(a generalization of the Majorana formula) reproduce the ground state energy of
the helium-like and lithium-like ions for $Z \leq 20$ in the domain of applicability
of NRQED, thus, at least, 3 s.d. It is noted that $\gtrsim 99.9\%$ of the
ground state energy is given by the variational energy
for properly optimized trial function of the form of (anti)-symmetrized product of
three (six) screened Coulomb orbitals for two-(three) electron system with 3 (7)
free parameters for $Z \leq 20$, respectively. It may imply that these trial
functions are, in fact,  {\it exact} wavefunctions in non-relativistic QED,
thus, the NRQED effective potential can be derived. It is shown that the sum
of relativistic and QED effects in leading approximation - 3 s.d. -
for both 2 and 3 electron systems is interpolated by 4th degree polynomial
in $Z$ for $Z \leq 20$.
\end{abstract}

\maketitle
\section*{Introduction}
\begin{figure}[tb]
  \begin{center}
   \includegraphics*[width=4.2in,angle=-90]{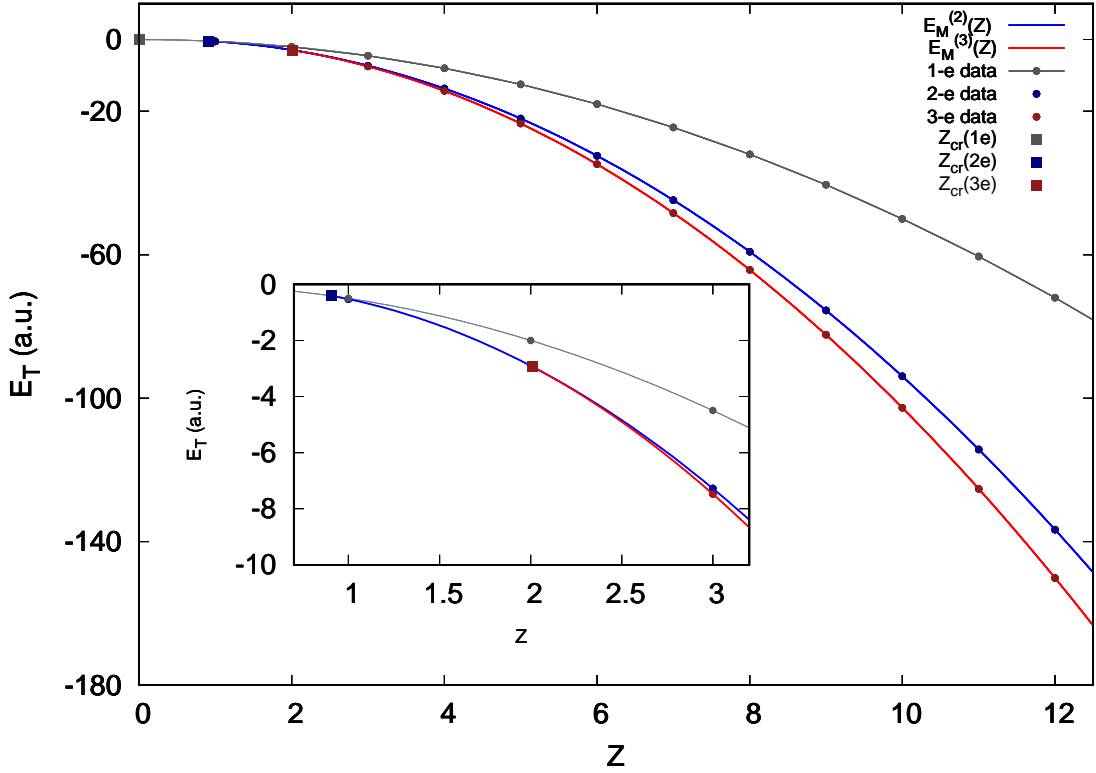}
    \caption{
    \label{Figure1}
    Ground state   energy {\it vs} $Z$ for one-, two-, three-electron atoms in
    NRQED in static approximation marked by bullets.
    Critical charges marked by filled squares. For plotted data see Tables
    \ref{Interpolation2e}, \ref{Interpolation3e} and for interpolating curves $E_M^{(2,3)} (Z)$
    see text at Section IV.
    }
  \end{center}
\end{figure}

We call {\it non-relativistic quantum electrodynamics} (NRQED) the
non-relativistic  quantum-mechanical theory of charged Coulomb particles
(without photons). In NRQED the Coulomb system of the $k$ electrons and an
infinitely-heavy, static, point-like charge $Z$: $(k\,e; Z)$ is described by a
Hamiltonian
\begin{equation}
\label{H}
  {\cal H}\ =\ -\frac{1}{2\,m} \sum_{i=1}^k \De_i \ -\ \sum_{i=1}^k \frac{Z}{r_i} \ +\
  \sum_{i>j=1}^k \frac{1}{r_{ij}}\ ,
\end{equation}
where $r_i$ is the distance from charge Z to $i$th electron of mass $m=1$ and
charge $e=-1$, $\De_i$ is three-dimensional Laplacian associated with $i$th
electron, $r_{ij}$ is the distance between the $i$th and $j$th electrons,
$\hbar=1$. Thus, energy is in atomic units (a.u.). It is widely known that
there exists a certain critical charge $Z_c$ above of which, $Z > Z_c$, the
system gets bound forming a $k$ electron atomic ion. It is also known that for
fixed $k$ the total energy of a bound state $E(Z)$, when exists, as the function
of integer charge $Z$ is very smooth, monotonously-decreasing negative function
which, with the growth of $Z$, is approaching to the sum of the energies of $k$
Hydrogenic ions, see for illustration Fig.\ref{Figure1} at $k=1,2,3$\,,
which behaves as $\sim k Z^2$ at large $Z$.

It is well known that domain of applicability of NRQED with static charge $Z$ is
limited due to finite-mass effects, as well as relativistic and QED effects,
and many other effects. The first three effects are dominant and define the
domain of applicability. For large $Z$ the relativistic effects become
significant and the Schr\"odinger equation should be replaced by the Dirac
equation. In order to remain in the Schr\"odinger equation formalism we limit our
consideration by $Z \leq 20$. Hence, the Schr\"odinger equation with Hamiltonian
(\ref{H}) describes NRQED within its domain of applicability. However,
non-relativistic energies $E$ emerging as eigenvalues of the Hamiltonian
(\ref{H}) are defined not only in the domain of applicability of NRQED but also
beyond of it, $E=E_{NRQED}+ \De E$. We call $\De E$ the "quantum"
correction to NRQED. Self-consistent solution of NRQED implies that $\De E$ is
of the same order of magnitude as the leading order of the sum of relativistic, QED and
finite mass corrections. Evidently, there are many solutions of NRQED
corresponding to different $\De E$ with different effective potentials other
than (\ref{H}). It seems natural to try to find an effective theory with
potential $V_{eff}$, other than (\ref{H}), leading to $E_{NRQED}$, which allows
the maximally simple exact solution. For $k=1$ the effective theory remains the
same NRQED (\ref{H}) since two-body Coulomb problem is exactly solvable with a
Coulomb orbital as the exact solution; as for $k=2$ it will be shown that the
celebrated Hylleraas function for the ground state \cite{Hylleraas} is one of
the simplest exact solutions of the effective theory, see below, which
reproduces $E_{NRQED}$. Needless to say that the knowledge of the exact
solutions allows us to find perturbatively the relativistic, QED and finite-mass
corrections as well as quantum correction.

In many applications, especially, in astrophysics and plasma physics we do not
need high accuracies, it might well be inside of the domain of applicability of
NRQED with static charges where relativistic, QED and mass effects can be
neglected. Hence, it is a natural problem to find such a domain of applicability
of NRQED explicitly for a few electron systems with static nuclear charge.
Surely, it may depend on the quantity we study. As the first step we consider
the ground state energy for 2-3 electron atomic systems which is predominantly
non-relativistic. We assume that everything is already known for one-electron,
hydrogen-like systems, see for review \cite{Eides}. As the second step we
show that the ground state energy is easily approximated, in particular, by
using the generalized Majorana formula but with coefficients different from
ones from 1/Z expansion, see below.

{\bf (A)}\, For two-electron case, $k=2$ (H${}^-$, He, Li${}^+$ etc) with
infinitely  heavy charge $Z$ (we call it the {\it static approximation})
the spectra of low-lying states was a subject of intense, sometimes
controversial, numerical studies (usually, each next calculation had found that
the previous one exaggerated its accuracy).
This program had run (almost) since the inception of quantum mechanics
\cite{Hylleraas} and had culminated at 2007 \cite{Nakashima:2007} when the
problem was solved for $Z = 1 - 10$ for the ground state with
overwhelmingly/excessively high accuracy ($\sim 35$ s.d.) from physical
point of view and it still continues.
Recently, it was checked that the energies found in \cite{Nakashima:2007}
are compatible with
$1/Z$-expansion up to 12 d.d. for $Z>1$ and 10 d.d. for $Z=1$,
see \cite{TL:2016}.
A time ago Nakashima-Nakatsuji made the impressive calculation of the ground
state energy of the 3-body problem $(2\,e; Z)$ with finite mass of nuclei
\cite{Nakashima:2008}. It was explicitly seen that taking into account the
finiteness of the nuclear mass changes in the   energy (taken in a.u.) the 4th
s.d.  for $Z=1,2$ and the 5th s.d.  for $Z=3 - 10$~. In present paper
using the Lagrange mesh method (see for basics \cite{Baye:2015}) we checked and
confirmed correctness of the 12 s.d. in both cases of infinite and finite
nuclear masses for $Z=1 - 10$ obtained in \cite{Nakashima:2007,Nakashima:2008}.
We also (re)calculated the ground state energies in both cases of infinite and
finite nuclear masses (in full geometry for the first time) for larger $Z=11,
12, 20, 30, 40, 50$ with accuracy of not less than 10 d.d. $^1$ \footnotetext[1]{Note that
for $Z=2$ the ground state energy difference for infinite and finite nuclear
mass (full geometry cases), see  Table I, coincides with the sum of the first
three orders in mass ratio found in \cite{Pachucki:2017} in 11 d.d.}

Note that in early days of quantum mechanics young Ettore Majorana in his
unpublished notes proposed an empirical one-parametric formula for the ground
state energy (in a.u.) versus nuclear charge $Z$:
\[
E=-Z^2 + \frac{5}{8} Z + b\ ,
\]
where $b$ is parameter, which can be found variationally, see, for historical
account, references and discussion \cite{E-S:2012}. At that time this formula
provided a reasonable description of energy; in principle, it can also permit
to make the calculation of the (first) critical charge $Z_c$, where
the ionization energy vanishes.

{\bf (B)}\, For three-electron case $k=3$ (Li, Be${}^+$ etc) accurate
calculations of the ground state energy for $Z=3 - 20$ were carried out in
\cite{Drake:1998} for both infinite and
finite nuclear masses. We believe that, at least, 10 s.d. obtained in these
calculations are confident. The effect of finiteness of the nuclear mass changes
4th - 3rd d.d. in the energy (in a.u.) when moving from small to large
$Z$.
For $Z = 15 - 20$ (and for infinite nuclear mass) the crosscheck of
compatibility of obtained results
with $1/Z$-expansion was also done: it was found that 5-6 d.d. in energy
coincide  \cite{Drake:1998}.
This coincidence provides us the confidence in the correctness of a certain
number of decimal digits (not less than six) which is sufficient for present
purposes. Note that finite mass effects were found perturbatively, taking into
account one-two terms in the expansion in electron-nuclei mass ratio.

Aim of the present paper is three-fold: (i) To localize the domain of
applicability of non-relativistic
QED for 2-3 electron systems with static nucleus,
(ii) to construct a simple interpolating function for the ground state
energy in full physics range of $Z \leq 20$ for $k=2,3$ which would reproduce
the ground state energy
with not less than twelve s.d. exactly, checking, in particular, applicability  of
the Majorana formula.
Such a number of exact s.d. is definitely inside of domain of applicability  of
non-relativistic QED with static nucleus, which is usually less than 4 d.d.;
(iii) to present a simple trial function for which the variational
energy describes the exact ground state energy in the domain of applicability
of non-relativistic QED with static nucleus with accuracy $99.9\%$,
thus, {\it de facto} it is the exact ground state NRQED wavefunction.

The paper is organized in the following way: Section I contains the analysis
of theoretical data defining the domain of applicability of NRQED, in Section
II the Puiseux expansion for small $Z$ and the Taylor expansion for large $Z$
for non-relativistic energies are presented; two-point interpolations are
introduced in Section III, Section IV is dedicated to Majorana formula and
its generalization, and a localization of critical charge, Section V is about
polynomial interpolation of the sum of leading relativistic and QED corrections,
Section VI presents the ``exact" wavefunctions for the ground state in NRQED
approximation for two and three electron ions and define
effective theory behind NRQED.

Throughout the paper all energies are given in atomic units (a.u.).
We use abbreviation {\it s.d.} for significant digits and {\it d.d.}
for decimal digits throughout the text.

\section{Domain of Applicability}

As the first step we have to collect data for the non-relativistic ground state
energies available  in literature for the cases of both infinite and
finite nuclear masses (taking the masses of the most stable nuclei, see
\cite{Audi:2003}) for two- and three-electron systems, see Table I, II,
respectively. In particular, this step is necessary to evaluate the effects
of finite nuclear mass to the ground state energy:
what leading significant (decimal) digit in energy is influenced by finite
mass effects.

{\bf (A)}\, For $k=2$ (two-electron ions) we collected in Table I the most
accurate data available in literature. As for the energies for
$Z=11, 12, 20, 30, 40, 50$ they were recalculated  employing the
Lagrange mesh method \cite{Baye:2015} for infinite mass case and calculated
for the first time for finite mass case (in full geometry), by using the
concrete  computer code designed for three-body studies
\cite{Hesse-Baye:2003,OT-PLA:2014}, where  details can be consulted.
In the past this method provided systematically the accuracy of 13-14 s.d.
for the ground state energy of various 3-body problems, see
e.g.~\cite{OT-PLA:2014}.
As for $Z=1 - 10$ the results {(rounded to 13-14 s.d.)} obtained in
\cite{Nakashima:2007,Nakashima:2008}
are also presented. All these energies were recalculated in the Lagrange  mesh
method and confirmed in {\it all} displayed digits at Table I.
Relativistic and QED effects (excluding $\log (\al Z)$ QED contributions)
in leading approximation were obtained for the first time
in \cite{Drake:1988} for $Z=1-100$, systematically they turned out negative;
they were recalculated in \cite{Yerokhin:2010} for $Z=2-12$.
Taking into account the smallest in powers of $\al$ terms only in
\cite{Yerokhin:2010} (excluding and including logarithmic contribution) in leading 3 s.d.
we collected them in Table I, columns 5 and 6.
In general, the difference between results of \cite{Drake:1988} and \cite{Yerokhin:2010}
occurred in one unit in the 3rd s.d. Eventually, we presented in column 5, Table I
the modern results by \cite{Yerokhin:2010} for $Z \leq 12$ and from \cite{Drake:1988} for $Z > 12$.
Taking polynomial interpolation in domain $Z=2-12$, see below Section V, where it reproduces
systematically all 3 s.d., we extrapolate it to $Z=1$ and $Z=20, 30, 40, 50$.
One can see that these effects leave unchanged the first 3-4 s.d. in the ground state energy.
It defines the domain of applicability of NRQED with static nuclei for $Z=1-20$ as 3-4 s.d.
in the ground state energy.
Let us emphasize that the special situation occurs for $Z=1$.
As the result of extrapolation from $Z \in [2,50]$ to $Z=1$ the sum of the leading corrections
$D_{rQED}^{(nl)} \equiv (E_{rel} + E_{QED}(nlog))$ (hence, excluding logarithmic terms) is
very small and positive, $2.85 \times 10^{-6}$\,a.u., while this sum in \cite{Drake:1988}
being of the same order of magnitude is negative $(-5.26 \times 10^{-6})$\,a.u.
It does {\it not} change our conclusion
about domain of applicability of NRQED. Present authors do not know a reason of this
discrepancy. To the best of our knowledge $D_{rQED}^{(l)} \equiv (E_{rel} + E_{QED})$
(hence, including logarithmic terms)
for $Z=1$ is not calculated. As for $Z=20$ the correction $D_{rQED}^{(l)}$
(which including logarithmic terms) in leading approximation was estimated recently
by Shabaev et al, \cite{MTS:2018}, see below.

{\bf (B)}\, For $k=3$ (three-electron ionic sequence) for infinite nuclear mass
the results by Yan et al, \cite{Drake:1998} are mostly presented in Table II, column 2.
Recently, for $Z=3,4$ these results were recalculated by Puchalski et
al, \cite{Pachucki:2008} using a more advanced variational method. This recalculation
confirmed 9 d.d. in energies obtained in \cite{Drake:1998}, but explicitly disagreed
in consequent 10-11th d.d. Since the results \cite{Pachucki:2008} give lower energies
than \cite{Drake:1998} we consider them as more accurate.
As for finite nuclear mass case for $Z=3-8$ the six d.d. only can be considered as
established, except for $Z=8$, see \cite{Drake:1998,Godefroid:2001,Pachucki:2008}.

In \cite{Drake:2003,Pachucki:2008} it was shown that for $Z=3$ the sum of
the leading relativistic ($E_{rel}$) and QED  ($E_{QED}$) corrections is of the same
order of magnitude as mass polarization. In particular, for $Z=3$ the sum of the
leading corrections $D_{rQED}\equiv E_{rel} + E_{QED} = -1.17\times 10^{-4}$ a.u.
while the mass correction $E_{mass}= 6.08\times 10^{-4}$ a.u., being of opposite sign,
see below Table II.  As for $Z=4$
the mass correction $E_{mass}= 8.99\times 10^{-4}$ a.u. \cite{Pachucki:2008} increases in about
$50\%$ with respect to $Z=3$, see Table II. Overall mass correction to ground state energy gives
contribution to 5th s.d. in the ground state energy for $Z=3,4$, see below, Table~II,~column~3.

We are unaware about the analysis of both QED and relativistic corrections for other values
of $Z$ of the same quality as in \cite{Drake:2003,Pachucki:2008} for $Z=3$.
Thus, we could only guess that for $Z=4-20$ these leading relativistic and QED corrections
leave unchanged 5-4-3 s.d. with growth of $Z$ up to $Z=20$ in the ground state energy
in static approximation (similarly to the two-electron case) $^2$ 
\footnotetext[2]{When this work was mostly completed, following our request the SPb University
group (A.V. Malyshev, I.I. Tupitsyn and V.M. Shabaev) kindly agreed
to make estimates of both relativistic and QED corrections in leading order for 3-electron ions
for different $Z$ \cite{MTS:2018}, see Table II, column 5.
We were informed that the method used is sufficiently accurate for large $Z$ but
it deteriorates at small $Z$. It turned out the estimates are consistent with results
found in \cite{Pachucki:2008} for $Z=3,4$ in $\sim 20\%$ and $\sim 1\%$, respectively.}.
Eventually, it defines the domain of applicability of non-relativistic QED for three electron atomic
systems with static nuclei as 3-4-5 s.d. in the total energy of ground state.

\section{Expansions}

It is well known since Hylleraas \cite{Hylleraas} that at large $Z$ the energy of $k$-electron
ion in static approximation admits the celebrated $1/Z$ expansion,
\begin{equation}
\label{1overZ}
  E(Z)\ =\ -B_0 Z^2 + B_1 Z + B_2 + O\bigg(\frac{1}{Z}\bigg)\ ,
\end{equation}
where $B_0$ is the sum of energies of $k$ Hydrogenic atoms, $B_1$ is the so-called electronic interaction energy,
which usually, can be calculated analytically. In atomic units $B_{0,1}$ are rational numbers.
In particular, for the ground state for $k=2$\,\cite{TL:2013},
\begin{equation}
\label{1overZ-2}
  B_0^{(2e)}\ =\ 1 \ ,\ B_1^{(2e)}\ =\ \frac{5}{8}\ ,\ B_2^{(2e)}\ =\,  -0.15766642946915\ ,
\end{equation}
and  $k=3$\,\cite{Drake:1998},
\begin{equation}
\label{1overZ-3}
  B_0^{(3e)}\ =\ \frac{9}{8} \ ,\ B_1^{(3e)}\ =\ \frac{5965}{5832} \ ,
  \ B_2^{(3e)}\ =\ -0.40816616526115\ ,
\end{equation}
respectively, where $B_2$ is the so-called electronic correlation energy. It was proved that
the expansion (\ref{1overZ}) for $k=2$ has a finite radius of convergence, see e.g. \cite{Kato:1980}.

In turn, at small $Z$, following the qualitative prediction by Stillinger and Stillinger \cite{Stillinger:1966} and
further quantitative studies performed in \cite{TG:2011}, \cite{TLO:2016}, there exists a certain value $Z_B > 0$
for which the non-relativistic ground state energy with static nuclei is given by the Puiseux expansion in a certain
fractional degrees
\begin{equation}
\label{PuiseuxGen}
\begin{split}
 E(Z) =& E_{B} + p_1 \left( Z-Z_{B} \right)
 + q_{{3}} \left( Z- Z_{B}\right)^{3/2} + p_{{2}} \left( Z - {\it Z_B} \right)^{2}
 +q_{{5}} \left( Z- Z_{B} \right) ^{5/2}
\\&
+ p_{{3}} \left( Z- Z_{B} \right)^{3}+q_{{7}} \left( Z- Z_{B} \right)^{7/2}
+ p_{{4}} \left( Z- Z_{B} \right)^{4} + \ldots \ ,
\end{split}
\end{equation}
where $E_{B}=E(Z_{B})$.
This expansion was derived numerically using highly accurate values of ground state energy in close vicinity
of $Z > Z_B$ obtained variationally. Three results should be mentioned in this respect for $k=2,3$:
(i) $Z_B$ is {\it not} necessarily equal to the critical charge, $Z_B \neq Z_c$,
(ii) the square-root term  $(Z- Z_{B})^{1/2}$ is absent, hence $E(Z)$ at $Z_B$ has square-root
branch point with exponent 3/2 and it may define the radius of convergence of $1/Z$ expansion (\ref{1overZ}), and,
(iii) seemingly the expansion (\ref{PuiseuxGen}) is convergent.
In particular, for the ground state at $k=2$\, the first coefficients in (\ref{PuiseuxGen}) are
\[
 Z_B^{(2e)}\ =\ {0.9048539992}\ ,\ E_B^{(2e)}\ =\ {-0.407932489} \ ,\
 p_1^{(2e)}\ =\, {-1.123475} \ ,
 \]
\begin{equation}
\label{k2par}
 \ q_3^{(2e)}\ =\, -0.197785\ ,\ p_2^{(2e)}\ =\, -0.752842\,,
\end{equation}
cf. \cite{TLO:2016}, while for $k=3$\,\cite{TLO:2016,TLOVN:2017},
\[
 Z_B^{(3e)}\ =\ 2.0090 \ ,\ E_B^{(3e)}\ =\ -2.934281\ ,\ p_1^{(3e)}\ =\ -3.390348\ ,\
 \]
\begin{equation}
\label{k3par}
 q_3^{(3e)}\ =\,  - 0.115425\,, p_2^{(3e)}\ = - 1.101372\,,
\end{equation}
respectively.

\section{Interpolation}

Let us introduce a new variable,
\begin{equation}
\label{lambdadef}
      {\la}^{2}=Z-{Z_B}\ .
\end{equation}
It can be easily verified that in $\la$ the expansion (\ref{PuiseuxGen}) becomes the Taylor expansion
while the expansion (\ref{1overZ}) is the Laurent expansion with the fourth order pole at $\la=\infty$.
The simplest interpolation matching these two expansion is given by a meromorphic function
\begin{equation}
\label{Int}
  -\,E_{N,4}(\la(Z))\ =\ \frac{P_{N+4}(\la)}{Q_N(\la)}\ \equiv\ \mbox{gPade}(N+4/N)_{n_0, n_{\infty}} (\la)
  \ ,
\end{equation}
which we call the {\it generalized, two point Pade approximant}. Here $P, Q$ are polynomials of degrees $N+4$ and $N$,
respectively
\[
   P_{N+4}=\sum_{\kappa=0}^{N+4} a_\kappa \la^\kappa\ ,\ Q_N=\sum_{\kappa= 0}^N b_\kappa \la^\kappa\ ,
\]
with normalization $Q(0)=1$, thus, $b_0=1$, the total number of free parameters in (\ref{Int})
is $(2N+5)$. It is clear that $P(0)=E_B$, thus $a_0 = E_B$.
The interpolation is made in two steps: (i) similarly to the Pade approximation theory some coefficients in (\ref{Int})
are found by reproducing exactly a certain number of terms $(n_0)$ in the expansion at small $\la$ and also a number
of terms $(n_{\infty})$ at large
$\la$-expansion, (ii) remaining undefined coefficients are found by fitting the numerical data, which we consider
as reliable, requiring the smallest
$\chi^2$. It is the state-of-the-art to choose $(n_0)$ and $(n_{\infty})$ appropriately.

For both cases $k=2,3$ in (\ref{Int}) we choose $N=5$, which is a minimal number
leading to correct 12 s.d., at least, in fit of exact ground state energy, see below.
It is assumed to reproduce {\it exactly} the first four terms in the Laurent expansion (\ref{1overZ}),
$n_{\infty}=4$, and the first three terms in the Puiseux expansion
(\ref{PuiseuxGen}), $n_{0}=3$. Thus, it leads us to the generalized, two point Pade Approximant
$\mbox{gPade}(9/5) (\la(Z))_{3,4}$. The remaining
eight free parameters in {Approximant}
\begin{equation}
\label{Int84}
    \mbox{gPade}(9/5) (\la)_{3,4}\ =\ \frac{E_B+a_1 \la+a_2 \la^2+a_3 \la^3+a_4 \la^4+a_5 \la^5+a_6 \la^6+a_7
    \la^7+a_8 \la^8 +a_9 \la^9}{1+b_1 \la+b_2 \la^2+b_3 \la^3+b_4 \la^4+b_5 \la^5}\ ,
\end{equation}
are found making fit. As for $k=2$, data from Table I, obtained by Nakashima-Nakatsuji \cite{Nakashima:2007} and
via the Lagrange mesh method \cite{OT-PLA:2014},
are fitted. While for $k=3$ data from Table II obtained by Yan et al \cite{Drake:1998} are used. In practice,
in order to construct the {Approximant} $\mbox{gPade}(9/5) (\la)_{3,4}$
we need to know the energies for eight values of $Z$ only.
In Table \ref{Paramsk23} the optimal parameters in $\mbox{gPade}(9/5) (\la)_{3,4}$ for $k=2,3$
are presented.
{Let us emphasize that in both cases the roots of denominator in $\mbox{gPade}(9/5) (\la)_{3,4}$ form complex-conjugated
pairs with mostly negative (or slightly positive) real parts!}

In general, expanding the function
$\mbox{gPade}(9/5) (\la(Z))$ with optimal parameters, see Table III, around $Z=Z_B{=0.9048539992}$ we get
\begin{equation*}
\begin{split}
  E^{(2e)}(Z) \simeq& {-0.40792489 - 1.123475\, (Z-Z_B) - 0.19599372613\, (Z- Z_B)^{3/2}  }
\\&
  {- 0.77430718831\, (Z-Z_B)^2  - 0.06604943902\,  (Z-Z_B)^{5/2} }+ \ldots\ ,
\end{split}
\end{equation*}
\begin{equation*}
\begin{split}
  E^{(3e)}(Z) \simeq&  - 2.934281 - 3.390348\, (Z-Z_B)
\\&
  -0.089012 \, (Z-Z_B)^{3/2}  - 1.253416 \, (Z-Z_B)^2 + 0.241838\, (Z-Z_B)^{5/2}\ldots\ ,
\end{split}
\end{equation*}
cf. (\ref{k2par})-(\ref{k3par}).

\vskip 1cm

\begin{sidewaystable}
\caption{\footnotesize
 Helium-like ions\,, the $1s^2\ {}^1S$ state energy (ground state):
 for $Z=1 \ldots 10$ \cite{Nakashima:2007} (note: for infinite nuclear mass case
 it coincides with $1/Z$ expansion in all displayed digits, see~\cite{TL:2016} and refs therein) and
 \cite{Nakashima:2008} (note: for finite nuclear mass case it coincides with Lagrange mesh results
 in all displayed digits, see text);
 for $Z=11,12$ (for infinite mass, column 2),~see \cite{TL:2016}, and Lagrange mesh results
 (present calculation, for both infinite and finite nuclear mass, column 3);
 for $Z= 20, 30, 40, 50$ the Lagrange mesh results presented for
 both infinite and finite nuclear mass cases (present calculation);
 for infinite nuclear mass case it is compared with fit (\ref{Int}) with $N=5$ (\ref{Int84})
 (column 7; for parameters see Table \ref{Paramsk23}).
 For infinite mass case (2nd column), {\it underlined} digits
 remain unchanged due to finite-mass effects (after its rounding),
 digits given by bold reproduced by fit (\ref{Int84}) ({before} rounding); Columns 5,6: the sum of leading relativistic
 and QED (excluding logarithmic term, column 5 and including, column 6) corrections (for infinite mass case) from \cite{Yerokhin:2010} given,
${}^{\dag}$ from \cite{Drake:1988},
 ${}^{\ddagger}$ the estimate \cite{MTS:2018},
 ${(\star)}$ the result of extrapolation from $Z \in [2-12]$, see (\ref{P4-He}), ${(\star\star)}$ extrapolation from (\ref{P4-Hev2}).
 Last column represents the variational energies from Anzatz (\ref{HeTrial}) (see text)
}
\begin{center}
\begin{tabular}{|c|r|r|r|c|c|c|c|c|}\hline
$Z$ & \multicolumn{3}{|c|}{$E$ (a.u.)}  &&  &Fit  (\ref{Int84}) & Fit ({\ref{Int40})}& Ansatz  \\
     &  \multicolumn{1}{|c}{Infinite mass} &\multicolumn{1}{c}{Finite mass} & \multicolumn{1}{c|}{Difference}
     & \multicolumn{1}{c|}{$D_{rQED}^{(nl)}$} & \multicolumn{1}{c|}{$D_{rQED}^{(l)}$} &   $\mbox{gPade}(9/5)_{3,4}$ & $\mbox{gPade}(4/0)_{2,1}$ &
(\ref{HeTrial}) \\
                \hline
 1    &  \underline{\bf -0.527}\,{\bf 751}\,{\bf 016 544} 4      &  -0.527\,445\,881\,1     &  $3.05\times 10^{-4}  $  & $ 2.85\times
10^{-6}$${(\star)}$ & $-1.68\times 10^{-5}$ ${(\star\star)}$ &   -0.527\,751\,016\, 54{8}  &   -0.5297   & -0.524   \\
 2    &  \underline{\bf -2.903}\,{\bf 724}\,{\bf 377 03}4        &  -2.903\,304\,557\,7     &  $4.20\times 10^{-4}  $  & $-1.12\times 10^{-4}$   &
$-8.17\times 10^{-5}$          &   -2.903\,724\,377\, 03{3 } &   -2.9049   & -2.900   \\
 3    &  \underline{\bf -7.279}\,{\bf 913}\,{\bf 412 66}9        &  -7.279\,321\,519\,8     &  $5.92\times 10^{-4}  $  & $-6.76\times 10^{-4}$   &
$-5.19\times 10^{-4}$          &   -7.279\,913\,412\, 6{65 } &   -7.2802   & -7.276   \\
 4    &  \underline{\bf -13.65}{\bf 5}\,{\bf 566}\,{\bf 238 4}2  &  -13.654\,709\,268\,2    &  $0.86\times 10^{-3}  $  & $-2.38\times 10^{-3}$   &
$-1.89\times 10^{-3}$          &  -13.655\,566\,238\, 4{1  } &  -13.6554   & -13.651  \\
 5    &  \underline{\bf -22.03}{\bf 0}\,{\bf 971}\,{\bf 580 2}4  &  -22.029\,846\,048\,8    &  $1.13\times 10^{-3}  $  & $-6.26\times 10^{-3}$   &
$-5.10\times 10^{-3}$          &  -22.030\,971\,580\, 2{3  } &  -22.0307   & -22.027  \\
 6    &  \underline{\bf -32.40}{\bf 6}\,{\bf 246\,601\, 90}      &  -32.404\,733\,488\,9    &  $0.15\times 10^{-2}  $  & $-1.37\times 10^{-2}$   & $-1.13\times 10^{-2}$          &  -32.406\,246\,601\, 90                  &  -32.4059   & -32.402  \\
 7    &  \underline{\bf -44.78}{\bf 1}\,{\bf 445}\,{\bf 148 7}7  &  -44.779\,658\,349\,4    &  $0.18\times 10^{-2}  $  & $-2.63\times 10^{-2}$   &
$-2.21\times 10^{-2}$          &  -44.781\,445\,148\, 7{ 5  } &  -44.7812   & -44.777  \\
 8    &  \underline{\bf -59.15}{\bf 6}\,{\bf 595}\,{\bf 122\,7}6 &  -59.154\,533\,122\,4    &  $0.21\times 10^{-2}  $  & $-4.61\times 10^{-2}$   &
$-3.93\times 10^{-2}$          &  -59.156\,595\,122\, 7{4  } &  -59.1565   & -59.152  \\
 9    &  \underline{\bf -75.53}{\bf 1}\,{\bf 712}\,{\bf 363\,9}6 &  -75.529\,499\,582\,5    &  $0.22\times 10^{-2}  $  & $-7.56\times 10^{-2}$   &
$-6.51\times 10^{-2}$          &  -75.531\,712\,363\, 9{3  } &  -75.5318   & -75.528  \\
10    &  \underline{\bf -93.90}{\bf 6}\,{\bf 806}\,{\bf 515\,0}4 &  -93.904\,195\,745\,9    &  $0.026\times 10^{-1} $  & $-1.17\times 10^{-1}$   &
$-1.02\times 10^{-1}$          &  -93.906\,806\,515\, {00  } &  -93.9071   & -93.903  \\
11    &  \underline{\bf -114.28}{\bf 1}\,{\bf 883}\,{\bf 776\,}1 & -114.279\,123\,929\,1    &  $0.028\times 10^{-1} $  & $-1.75\times 10^{-1}$   &
$-1.53\times 10^{-1}$          & -114.281\,883\,776\, {0   } & -114.2824   & -114.278 \\
12    &  \underline{\bf -136.65}{\bf 6}\,{\bf 948}\,{\bf 312\,}7 & -136.653\,788\,023\,4    &  $0.032\times 10^{-1} $  & $-2.50\times 10^{-1}$   &
$-2.20\times 10^{-1}$          & -136.656\,948\,312\, {6   } & -136.6577   & -136.653 \\[5pt]
\hline
\end{tabular}
\end{center}
\label{Interpolation2e}
\end{sidewaystable}%

\addtocounter{table}{-1}
\begin{sidewaystable}
\caption{\footnotesize
(continuation)}
\begin{center}
\begin{tabular}{|c|r|r|r|c|c|c|c|c|}\hline
$Z$ & \multicolumn{3}{|c|}{$E$ (a.u.)}  &&  &Fit  (\ref{Int84}) & Fit ({\ref{Int40})}& Ansatz  \\
    &  \multicolumn{1}{|c}{Infinite mass} &\multicolumn{1}{c}{Finite mass} & \multicolumn{1}{c|}{Difference}
    & \multicolumn{1}{c|}{$D_{rQED}^{(nl)}$}  & \multicolumn{1}{c|}{$D_{rQED}^{(l)}$} &   $\mbox{gPade}(9/5)_{3,4}$ & $\mbox{gPade}(4/0)_{2,1}$ &
(\ref{HeTrial}) \\
                \hline
20  &  \underline{\bf -387.65}{\bf 7}\,{\bf 233}\,{\bf 83}3 2      & -387.651\,875\,961\,4    & $5.36\times 10^{-3}$  & $-2.00^{\dag} $   & $-1.84^\ddagger$  & -387.657\,233\,834\,0 & -387.6604   & -387.653  \\
    &                                                        &                          &                       & $-2.05$ ${(\star)}$
                  & $-1.85 {(\star\star)}$ &
      &             &    \\
30  &  \underline{\bf -881.40}{\bf 7}\,{\bf 377}\,{\bf 4}88 3      & -881.399\,778\,896\,1    & $7.60\times 10^{-3}$  & $-10.46^{\dag} $ &  &
-881.407\,377\,492\,{6} & -881.4142   & -881.403  \\
    &                                                        &                          &                       & $-10.66$ ${(\star)}$ & $-9.78 {(\star\star)}$&
              &             &   \\
40  &  \underline{\bf -1\,575.15}{\bf 7}\,{\bf 449}\,{\bf 5}25 6   & -1575.147\,804\,148\,0   & $9.65\times 10^{-3}$  &  $-33.78^{\dag}$ &  &
-1\,575.157\,449\,53{5} & -1575.1684  & -1575.153 \\
    &                                                        &                          &                       & $-34.19$ ${(\star)}$ & $-31.62 {(\star\star)}$ &
     &             &   \\
50  &  \underline{\bf -2\,468.90}{\bf 7}\,{\bf 492}\,{\bf 8}12\,7   & -2468.895\,972\,259\,1   & $1.15\times 10^{-2}$  & $-84.13^{\dag}$  &   &
-2\,468.907\,492\,82{8} & -2468.9230  & -2468.903 \\
    &                                                        &                          &                       & $-84.20$ ${(\star)}$  & $-78.26 {(\star\star)}$&
              &             &   \\
\hline
\end{tabular}
\end{center}
\label{Interpolation2e-1}
\end{sidewaystable}%

\renewcommand{\arraystretch}{0.7} 

\begin{sidewaystable}
\caption{\footnotesize
 Lithium-like ions\,, the lowest, $1s^2\,2s\ {}^2 S$ state energy:
 for $Z=3 - 20$ \cite{Drake:1998}  (infinite and finite nuclear mass cases);
 it is compared with the fit (\ref{Int}) with $N=5$ (\ref{Int84}).
 For $Z=3,4$ energies for (in)finite mass cases
 displayed in 1st row from \cite{Drake:1998}, in 2nd row  marked ${}^{(\dag)}$ from \cite{Pachucki:2008}.
 For $Z=3\ldots 8$ finite mass results displayed in 3-2 rows are from \cite{Godefroid:2001}
 with the absolute difference calculated with respect to the infinite mass case  \cite{Drake:1998};
 for infinite nuclear mass case it is compared with fit (\ref{Int84}), column 6
 (for parameters see Table \ref{Paramsk23}).\\
 For infinite mass case (column 2), underlined digits remain unchanged due to finite-mass effects (after its rounding),
 digits in bold reproduced by fit (\ref{Int84}) (after rounding). Last column represents the variational energies obtained with Anzatz (\ref{GStrialfunct})-(\ref{chi}).\\
 Leading relativistic+QED corrections (for infinite mass case) $D_{rQED}$ for $Z=3$ marked ${}^{(\dag)}$ from \cite{Drake:2003,Pachucki:2008}, marked
by ${}^{(\star)}$ from \cite{MTS:2018}, see also [36];
 for $Z \geq 4$ the estimates for leading relativistic+QED corrections
 (for infinite mass case), column 5, from \cite{MTS:2018}.
}


\begin{center}{\footnotesize
\begin{tabular}{|c|r|r|r|c|c|c|c|c|}\hline
$Z$ & \multicolumn{3}{|c|}{$E$ (a.u.)} & & Interpolation &Fit  (\ref{Int84}) &  Fit (\ref{Int40}) & Ansatz \\
    &  \multicolumn{1}{|c}{Infinite mass} &\multicolumn{1}{c}{Finite mass} & \multicolumn{1}{c|}{Difference} & $D_{rQED}$ & (\ref{P4-Li-t})& $\mbox{gPade}(9/5)_{3,4}$    & $\mbox{gPade}(4/0)_{2,1}$ & (\ref{GStrialfunct})-(\ref{chi})  \\
 \hline
     3              &\  \underline{\bf -7.47}{\bf 8\,060\,323\,65}0               &    -7.477\, 451\, 884\, 70  & $ 6.08\times 10^{-4}$  & $
-5.2\times 10^{-4}$ ${}^{(\star)}$ &                         & -7.478\,060\,323\,651  & -7.495   & -7.455 \\
     ${}^{(\dag)}$   &\  \underline{\bf -7.47}{\bf 8\,060\,323}\,91                &    -7.477\, 452\, 121\, 22  & $ 6.08\times 10^{-4}$  & $-1.17\times 10^{-4}$                & $-1.17\times 10^{-4}$   &                  &          &   \\
                    &                                                             &    -7.477\, 452\, 048\, 02  & $ 6.08\times 10^{-4}$  &                                      &                         &                  &          &  \\
     4              &\qquad \ \underline{\bf -14.32}{\bf 4\,7}{\bf 63\,176\,465}   &   -14.323\, 863\, 441\, 3   & $ 9.00\times 10^{-4}$  &
$-1.96\times 10^{-3}$                & $-1.25\times 10^{-3}$   & -14.324\,763\,176\,47  & -14.340  &  -14.271\\
    ${}^{(\dag)}$    &\qquad \ \underline{\bf -14.32}{\bf 4\,7}{\bf 63\,176}\,78   &   -14.323\, 863\, 713\, 6   & $ 8.99\times 10^{-4}$  &                                      &                         &                  &          &  \\
                    &                                                             &   -14.323\, 863\, 687\, 1   & $ 8.99\times 10^{-4}$  &                                      &                         &                  &          &  \\
     5              &  \underline{\bf -23.42}{\bf 4\,6}{\bf 05\,720\,96}           &   -23.423\, 408\, 020\, 3   & $ 1.20\times 10^{-3}$  &
$-5.38\times 10^{-3}$                & $-4.63\times 10^{-3}$   & -23.424\,605\,720\,96  & -23.436  &  -23.330\\
                    &                                                             &   -23.423\, 408\, 350\, 5   & $ 1.20\times 10^{-3}$  &                                      &                         &                  &          &  \\
     6              &  \underline{\bf -34.77}{\bf 5\,5}{\bf 11\,275\,6}3           &   -34.773\, 886\, 337\, 7   & $ 1.62\times 10^{-3}$  &
$-1.21\times 10^{-2}$                & $-1.15\times 10^{-2}$   & -34.775\,511\,275\,63  & -34.782  &  -34.633\\
                    &                                                             &   -34.773\, 886\, 826\, 3   & $ 1.62\times 10^{-3}$  &                                      &                         &                  &          &  \\
     7              &  \underline{\bf -48.37}{\bf 6\,8}{\bf 98\,319\,1}4           &   -48.374\, 966\, 777\, 1   & $ 1.93\times 10^{-3}$  &
$-2.39\times 10^{-2}$                & $-2.35\times 10^{-2}$   & -48.376\,898\,319\,12  & -48.380  &  -48.182 \\
                    &                                                             &   -48.374\, 967\, 352\, 1   & $ 1.93\times 10^{-3}$  &                                      &                         &                  &          &  \\
     8              &  \underline{\bf -64.22}{\bf 8\,5}{\bf 42\,082\,7}0           &   -64.226\, 301\, 948\, 5   & $ 2.24\times 10^{-3}$  &
$-4.29\times 10^{-2}$                & $-4.26\times 10^{-2}$   & -64.228\,542\,082\,71  & -64.229  &  -63.967\\
                    &                                                             &   -64.226\, 375\, 998\, 3   & $ 2.17\times 10^{-3}$  &                                      &                         &                  &          &  \\
     9              &  \underline{\bf -82.33}{\bf 0\,3}{\bf 38\,097\,3}0           &   -82.327\, 924\, 832\, 7   & $ 2.41\times 10^{-3}$  &
$-7.13\times 10^{-2}$                & $-7.12\times 10^{-2}$   & -82.330\,338\,097\,35  & -82.328  &  -81.993 \\
    10              &  \underline{\bf -102.68}{\bf 2}\,{\bf 231\,482\,4}          &  -102.679\, 375\, 319\,     & $ 2.86\times 10^{-3}$  & $-1.12\times 10^{-1}$                & $-1.12\times 10^{-1}$   & -102.682\,231\,482\,4  & -102.678 &  -102.243\\
    11              &  \underline{\bf -125.28}{\bf 4\,1}{\bf 90\,753\,6}          &  -125.281\, 163\, 823\,     & $ 3.03\times 10^{-3}$  & $-1.69\times 10^{-1}$                & $-1.69\times 10^{-1}$   & -125.284\,190\,753\,6  & -125.279 &  -124.730\\
    12              &  \underline{\bf -150.13}{\bf 6\,1}{\bf 96\,604\,5}          &  -150.132\, 723\, 126\,     & $ 3.47\times 10^{-3}$  & $-2.44\times 10^{-1}$                & $-2.44\times 10^{-1}$   & -150.136\,196\,604\,5  & -150.131 &  -149.440\\
    13              &  \underline{\bf -177.23}{\bf 8\,2}{\bf 36\,560\,0}          &  -177.234\, 594\, 529\,     & $ 3.64\times 10^{-3}$  & $-3.43\times 10^{-1}$                & $-3.43\times 10^{-1}$   & -177.238\,236\,560\,0  & -177.233 &  -176.315\\
    14              &  \underline{\bf -206.59}{\bf 0}\,{\bf 302\,212\,3}          &  -206.586\, 211\, 017\,     & $ 4.09\times 10^{-3}$  & $-4.69\times 10^{-1}$                & $-4.69\times 10^{-1}$   & -206.590\,302\,212\,3  & -206.585 &  -205.507\\
    15              &  \underline{\bf -238.19}{\bf 2\,3}{\bf 87\,694}\,1          &  -238.188\, 129\, 642\,     & $ 4.26\times 10^{-3}$  & $-6.27\times 10^{-1}$                & $-6.27\times 10^{-1}$   & -238.192\,387\,694\,2  & -238.188 &  \\
    16              &  \underline{\bf -272.04}{\bf 4\,488\,790\,1}                   &  -272.039\, 780\, 017\,     & $ 4.71\times 10^{-3}$  & $-8.23\times 10^{-1}$                & $-8.23\times 10^{-1}$   & -272.044\,488\,790\,1  & -272.042 &  \\
    17              &  \underline{\bf -308.14}{\bf 6}\,{\bf 602\,395\,3}          &  -308.141\, 728\, 192\,     & $ 4.87\times 10^{-3}$  &  -1.06                              &  -1.06                 & -308.146\,602\,395\,3  & -308.146 &  \\
    18              &  \underline{\bf -346.49}{\bf 8}\,{\bf 726\,173\,7}          &  -346.493\, 932\, 364\,     & $ 4.79\times 10^{-3}$  &  -1.35                              &  -1.35                 & -346.498\,726\,173\,7  & -346.500 &  \\
    19              &  \underline{\bf -387.10}{\bf 0}\,{\bf 858\,334\,6}          &  -387.095\, 367\, 736\,     & $ 5.49\times 10^{-3}$  &  -1.69                              &  -1.69                 & -387.100\,858\,334\,6  & -387.105 &  \\
    20              &  \underline{\bf -429.95}{\bf 2\,9}{\bf 97\,482\,8}          &  -429.947\, 053\, 487\,     & $ 5.94\times 10^{-3}$  &  -2.10                              &  -2.10                 & -429.952\,997\,482\,8  & -429.961 &  \\
 \hline
\end{tabular}
}
\end{center}
\label{Interpolation3e}
\end{sidewaystable}%

\begin{table}
\caption{Parameters in $\mbox{gPade}(9/5)_{3,4}\,(\la(Z))$ for $k=2,3$ rounded to 12 s.d.,
  3 constraints imposed for the small $\la$ and 4 constraints for the large $\la$ limits,
  see (\ref{Int84}).
  For $k=2$ the fit done for data from Table \ref{Interpolation2e}, column 2
  corresponding to $Z=1, \ldots 10$.  For $k=3$ the fit done for data from Table \ref{Interpolation3e}, column 2 corresponding to $Z=3,  \ldots 20$.}
\begin{center}
\begin{tabular}{|c|c|c|}\hline
parameters & $k=2$  &  $k=3$   \\
\hline
         $a_{0}$ &\  {     -0.40792489    } \ &   \  -2.934281         \\
         $a_{1}$ &\  {     -1.7720945736  } \ &   \  -4.985319466822   \\
         $a_{2}$ &\  {     -5.09227274076 } \ &   \  -12.51820588130   \\
         $a_{3}$ &\  {    -11.4229111014 } \ &   \  -14.24645657801    \\
         $a_{4}$ &\  {    -16.6866531073 } \ &   \  -15.83837203461    \\
         $a_{5}$ &\  {    -28.2878912031 } \ &   \  -15.74077387390    \\
         $a_{6}$ &\  {    -21.7247609889 } \ &   \  -8.13280408452     \\
         $a_{7}$ &\  {    -31.4514043213 } \ &   \  -7.836711216798    \\
         $a_{8}$ &\  {    -10.1253004761 } \ &   \  -1.490328878341    \\
         $a_{9}$ &\  {    -13.4157573972 } \ &   \  -1.485186602328    \\
         $b_{0}$ &\  {     1.0          } \ &   \   1.0                \\
         $b_{1}$ &\  {     4.34416878449} \ &   \   1.698991837122     \\
         $b_{2}$ &\  {     9.72923652871} \ &   \   3.110764743151     \\
         $b_{3}$ &\  {    15.5576492283 } \ &   \   2.861781316187     \\
         $b_{4}$ &\  {    10.1253004761 } \ &   \   1.324736780747     \\
         $b_{5}$ &\  {    13.4157573972 } \ &   \   1.320165868736     \\
          \hline
\end{tabular}
\end{center}
\label{Paramsk23}
\end{table}%

\begin{table}
\caption{Parameters in $\mbox{gPade}(4/0)_{2,1}\,(\la(Z))$ for $k=2,3$, see (\ref{Int40}),  found by
         fitting the same data as in Table \ref{Paramsk23}}
\begin{center}
\begin{tabular}{|c|c|c|}\hline
parameters & $k=2$  &  $k=3$   \\
\hline
         $a_{0}$ &\  -0.407924         &\   -2.934281     \\
         $a_{1}$ &\   0.0              &\    0.0          \\
         $a_{2}$ &\  -1.184891         &\   -3.485218     \\
         $a_{3}$ &\  -0.000027         &\   -0.002469     \\
         $a_{4}$ &\  -1.0              &\   -9/8          \\
           \hline
\end{tabular}
\end{center}
\label{Paramsk23min}
\end{table}

In Table I and II the results of interpolations for $k=2$ and $k=3$ are presented, respectively.
In general, {for $k=2,3$ the difference in energy occurs systematically in 13rd s.d. or, sometimes, in 14th
s.d. for all range of $Z$ studied, except for $Z=40,50$ for $k=2$, where it occurs in 12th s.d.}

The analysis of relativistic and QED corrections for two-electron system performed for $Z=2-12$ in
\cite{Yerokhin:2010,Pachucki:2017}, see Table I, shows that they are small or comparable with respect
to the mass polarization effects for $Z=1,2,3$ and then become larger (and dominant) for $Z > 3$.
Note that the relativistic and QED corrections systematically are of opposite sign.

Similar analysis of relativistic and QED corrections of three-electron system, performed for $Z=3$
in \cite{Drake:2003,Pachucki:2008}, shows that they contribute to the 1st s.d. in the energy difference between
infinite and finite mass cases. For both cases of
2- and 3-electron systems the question about the order of relativistic and QED effects for large $Z$
needs to be investigated. We can only guess that
for both systems the domain of applicability of static approximation for any $Z$ is limited by 3 s.d.
in the ground state energy.

Interestingly, the simplest interpolation $\mbox{gPade}(4/0)_{2,1}\,(\la(Z))$ (which is in fact the
terminated Puiseux expansion) with two fitted parameters $a_{2,3}$,
\begin{equation}
\label{Int40}
    \mbox{gPade}(4/0) (\la)_{2,1}\ =\ E_B + a_1 \la + a_2 \la^2 + a_3 \la^3 + a_4 \la^4\ ,
\end{equation}
with other parameters taken from Table \ref{Paramsk23min}  $^3$ \footnotetext[3]{Note that the parameter $a_3$ is small, hence, the term $a_3 \la^3$ does not play a significant role in description of energy dependence at $Z \leq 20$} and $a_1=0$ reproduces 3-4 s.d. in ground state energy in static approximation for both systems in physics range of $Z \leq 20$,
see~Tables~I,II\,. These 3-4 s.d. remain unchanged if all finite-mass, relativistic and QED effects are taken into account. It implies the exact reproduction of the domain of applicability of non-relativistic QED in static approximation for the ground state total energy!

\section{Majorana Formula and the (first) critical charge}

Originally, the Majorana formula $^4$ \footnotetext[4]{Recently, this formula was found, in fact,
in unpublished and unknown till recently notes by E Majorana circa 1930, see e.g. Eq.(22) in \cite{E-S:2012} and references therein}
\begin{equation}
\label{majorana}
  E^{(2)}_M(Z)\ =\ -e_2 Z^2 + e_1 Z + e_0\ ,\quad e_2 = 1\ ,
\end{equation}
was written for the ground state energy of two-electron system in static approximation as the variational
energy for the trial function $\sim \exp{\{-Z(r_1+r_2)\}}$\ (i), then it was re-interpreted as the first
three terms of $1/Z$-expansion, cf. (\ref{1overZ}) (ii) and then the parameter $e_0$ was set free to be chosen to get the best description of experimental data for Helium atom, $Z=2$, \cite{E-S:2012} (iii)  $^5$
\footnotetext[5]{For one-electron case, $k=1$, the Majorana formula at $e_2 = 1/2$ and $e_{1,0}=0$ is exact. In fact, $E^{(2)}_M(Z)=\mbox{gPade}(4/0)  (\la)_{0,2}$ at $a_1=0$}.
None of these three considerations (i)-(iii) had led to accurate description of data being limited to 1-2 s.d. and, sometimes, to 3 s.d., see Table \ref{Zc2}. Note since E Majorana was likely the first who treated $Z$ as continuous parameter, those three considerations provided for the first time the values of the (first) critical charge $Z_c$, where the one-electron ionization energy vanishes
$I(Z_c)=0$, see Table \ref{Zc2}.
The situation changes dramatically if parameter $e_{0}$ or two parameters $e_{1,0}$ are varied to get the best fit of data in the whole domain $Z \in [1,50]$: the Majorana formula reproduces consistently, at least, 3-4 s.d. in the ground state energy leading to the exact NRQED energies at $Z \in [1,20]$ in a way similar to (\ref{Int40}) (with parameters from Table \ref{Paramsk23min})  
$^6$ 
\footnotetext[6]{Note that although fitted parameters $e_{1,0}$ are very close to the second-third coefficients in $1/Z$-expansion for two-electron systems (\ref{1overZ-2}), 
but different. Namely, due to the difference we are able to get high quality approximation}.
In this case the Majorana formula (\ref{majorana}) gives rather accurate value of the first critical charge $Z_c^{(2)}$:
\[
    I^{(2)}(Z^{(2)}_c)\ =\ E^{(2)}_M(Z^{(2)}_c) + \frac{(Z^{(2)}_c)^2}{2}\ =\ 0\ ,
\]
see Table \ref{Zc2} (3rd line), which is in good agreement with the value of the critical charge predicted
by the Approximants given by (\ref{Int40}) or by (\ref{Int84}) as well as with the exact result
\cite{Drake:2014, OT-PLA:2014} $^7$
{\footnotetext[7]{{Energy at $Z_2^{(2)}$ found in \cite{Drake:2014, OT-PLA:2014} is reproduced by Approximant (\ref{Int84}) in 10 s.d.(!), the energy difference is $\sim 1 \times 10^{-10}$, cf. Table I}}}.
Eventually, the Majorana formula with fitted coefficients $e_{1,0}$ can be considered as accurate interpolation of the ground state energy curve for two-electron system on Fig. \ref{Figure1}. Note that $1/Z$-expansion can be constructed for any excited state and it always has the form (\ref{1overZ}). By taking a linear superposition of the first three terms we will arrive at the Majorana type formula (\ref{majorana}).
By keeping the coefficient $e_2$ equal to one found in $1/Z$-expansion and fitting the coefficients $e_{1,0}$
we should get a reasonable description of $Z$-dependence of the energy of an excited state.
It will be checked elsewhere.
\begin{table}[htp]
\caption{\label{Zc2} Helium-like system:
      relative deviation $\delta_E=|\frac{E-E_{exact}}{E_{exact}}|$ {\it vs} $Z$ for Majorana and generalized
      Pade approximations, the first critical charge $Z^{(2)}_c$: $I^{(2)}(Z^{(2)}_c)=0$, see text. }
  \begin{center}
\begin{tabular}{|l|ll|c|c|}
\hline
Approximation &  \multicolumn{2}{|c|}{ parameters}  & $\de_E$ & $Z^{(2)}_c$ \\
&&&$(Z = 1 - 50)$ &\\
\hline
\multirow{3}{*}{Majorana formula (\ref{majorana}) }
& \ $e_1=\frac{5}{8}$        &\hspace{-80pt} $e_0= -\frac{25}{256}$ \    &$\lesssim 0.1    $ &\ \ 1.066 942
\phantom{\huge I} \\[5pt]
& \ $e_1=\frac{5}{8}$        &\hspace{-80pt}$e_{0,\rm fit}= -0.155573$\ &$\lesssim 0.005  $ &\ \ 0.906 982
\phantom{I}  \\[5pt]
& \ $e_{1,\rm fit}=0.624583$ &\hspace{-80pt}$e_{0,\rm fit}=-0.153282$ \ &$\lesssim 0.0002 $ &\ \ 0.913 617
\phantom{\Large I}    \\
\hline
\multirow{2}{*}{Generalized Pade}
& \ ${\rm gPade}(4/0)_{2,1}$ (\ref{Int40})\hspace{20pt} (see Table \ref{Paramsk23min} ) & &$\lesssim 0.003  $ & 0.910 007   \\
& \ ${\rm gPade}(9/5)_{3,4}$ (\ref{Int84})\hspace{20pt} (see Table \ref{Paramsk23} ) &  &$\lesssim 10^{-12} $ & {0.911 028 22} \\
\hline
Exact \cite{Drake:2014, OT-PLA:2014} (rounded) & &&&  0.911 028 22  \\
\hline
\end{tabular}
\end{center}
\end{table}%

It is natural to try to explore the Majorana formula (\ref{majorana}) to describe the $Z$-dependence of the ground state energy of three-electron system, $E^{(3)}_M(Z)$, with fixed $e_2=9/8$ and fitting $e_{1,0}$.
Straightforward fit shows that the Majorana formula reproduces at least 3-4 s.d. in the ground state energy leading to the exact NRQED energies at $Z \in [3,20]$ in a way similar to (\ref{Int40}) (with parameters from Table \ref{Paramsk23min}) $^8$
\footnotetext[8]{Note that although fitted parameters $e_{1,0}$ are very close, but are slightly different, to the second-third coefficients in $1/Z$-expansion for three-electron systems (\ref{1overZ-3}). Namely, due to the difference we are able to get high quality approximation}.
In this case the Majorana formula (\ref{majorana}) gives a reasonable value of the first critical charge $Z^{(3)}_c$
\[
    I^{(3)}(Z^{(3)}_c)\ =\ E^{(3)}_M(Z^{(3)}_c) - E^{(2)}_M(Z^{(3)}_c)\ =\ 0\ ,
\]
see Table \ref{Zc3}. For the case of generalized Pade approximations (\ref{Int40}) and (\ref{Int84})
the predicted first critical charge coincides with the second critical charge \cite{TG:2011}:
it is the minimal nuclear charge for ground state function becomes non-normalizable.

\begin{table}[htp]
\caption{\label{Zc3}
      Lithium-like system:
      relative deviation $\delta_E=|\frac{E-E_{exact}}{E_{exact}}|$ {\it vs} $Z$ for Majorana and generalized Pade approximations, the first critical
charge $Z_c$: $I^{(3)}(Z_c)=0$, see text }
  \begin{center}
\begin{tabular}{|l|ll|c|c|}
\hline
Approximation &  \multicolumn{2}{|c|}{parameters}  & $\de_E$ & $Z^{(3)}_c$ \\
&&&$(Z = 3 - 20)$ &\\
\hline
   Majorana formula (\ref{majorana})  &\ $e_2=9/8$ \quad  $e_{1,fit}=1.023260$ & $e_{0,fit}=-0.416432$ \ &\ $\lesssim 0.0009$\ &\ 2.256 \\[5pt]
\hline
\multirow{2}{*}{Generalized Pade}
&\ ${\rm gPade}(4/0)_{2,1}$ (\ref{Int40}) (see Table \ref{Paramsk23min} )\ & &\ $\lesssim 0.002$ \ &\   2.009   \\
&\ ${\rm gPade}(9/5)_{3,4}$ (\ref{Int84}) (see Table \ref{Paramsk23} )\ & &$\ \lesssim 10^{-11}$\ &\   2.009   \\
\hline
\end{tabular}
\end{center}
\label{default}
\end{table}%
Eventually, one can state that the Majorana formula with fitted coefficients $e_{1,0}$ can be considered as accurate interpolation of the ground state energy curve for three-electron system in NRQED approximation, see Fig.\ref{Figure1}.

Apparently, $1/Z$-expansion can be constructed for any excited state of three electron system and it
has the form (\ref{1overZ}). By taking a linear superposition of the first three terms we will arrive
at the Majorana type formula (\ref{majorana}).
Seemingly, by fitting the coefficients $e_{1,0}$ with $e_2$ taken from $1/Z$-expansion we should get
a reasonable description of $Z$-dependence of the energy of an excited state of $(3e;Z)$ system.
It will be checked elsewhere.

It seems important to try to find $Z$-dependent trial function, if exists, which would lead
to the Majorana formula (\ref{majorana}) with coefficients from Table \ref{Zc2} (3rd line) and
from Table \ref{Zc3} (1st line) for 2-3 electron cases, respectively.

\section{Approximating the sum of leading relativistic and QED corrections {\it versus}\ $Z$}

In previous Section it was shown that the second degree polynomial in $Z$ (Majorana formula (\ref{majorana}))
approximates accurately NRQED ground state energies for Helium like and Lithium like systems.
We conjectured that it remains true for the energy of any excited state and for other atom-like systems
with $k > 3$. It is interesting to try to approximate the sum of leading relativistic and QED corrections
$D_{rQED}=D_{r}+D_{QED}$ of orders $\al^4$ and $\al^5$, respectively, (excluding and including logarithmic
contributions); see e.g. \cite{Drake:1988,Yerokhin:2010}, presented in columns 5 and 6 in Tables I (and II)
for different $Z$. It does not look as a simple task since from $Z=1\,(3 \ \mbox{for}\ k=3)$ to
$Z=50\, (20 \ \mbox{for}\  k=3)$ it changes in six orders of magnitude(!), see Table I $(II \ \mbox{for}\ k=3)$.

Following standard formulas for the sum $D_{rQED}$ in leading order, see e.g. \cite{Drake:1988},
at large $Z$,  it behaves like $\sim Z^4$, if $\log Z$-dependence is neglected.
Hence, as interpolating function we choose naively a polynomial in $Z$ of degree 4,
\begin{equation}
\label{P4}
   D_{rQED}\ =\ a_0\ +\ a_1\, Z\ +\ a_2\, Z^2\ +\ a_3\, Z^3\ +\ a_{4}\,Z^4\ .
\end{equation}
For the Helium-like sequence we choose 11 integer points $Z \in [2,12]$ and make fit of 5 parameters
$a_{0,1,2,3,4}$ in (\ref{P4}) with goal to reproduce all 3 s.d. in $D_{rQED}$
(excluding $D_{rQED}^{(nl)}$ and including logarithmic contributions $D_{rQED}^{(l)}$)
printed in Table I, column 5 and 6.
This goal is achieved for both $D_{rQED}^{(nl)}$ $^9$ \footnotetext[9]{Except for $Z=3, 8$,
where for unclear reason the deviation in one unit in the 3rd s.d. was seen},
where we predict $D_{rQED}^{(nl)}$ at $Z=1$,
and for total $D_{rQED}^{(l)}$.
For larger $Z=20,30,40,50$ the value of $D_{rQED}^{(nl)}$ is described with relative accuracy
$\sim 10^{-3}$, see Table \ref{Interpolation2e} (continuation). Final expression for interpolating
polynomial $D_{rQED}^{(nl)}$ reads,
\begin{equation}
\label{P4-He}
 D_{rQED}^{nl,\rm He-like} = (2.245 - 0.6938\, Z - 2.311\, Z^2 + 2.440\, Z^3 - 1.395\, Z^4)\times 10^{-5}\ ,
\end{equation}
while for the total relativistic and QED correction in leading approximation it is,
\begin{equation}
\label{P4-Hev2}
 D_{rQED}^{l,\rm He-like} = (-7.174 + 11.046\, Z - 7.976\, Z^2 + 3.749\, Z^3 - 1.324\, Z^4)\times 10^{-5}\ .
\end{equation}

For Lithium-like sequence we choose 9 integer points $Z \in [5,13]$ and make fit of 5 parameters
$a_{0,1,2,3,4}$ in (\ref{P4}) with the first goal to reproduce all three s.d. in estimates of $D_{rQED}$
due to \cite{MTS:2018}, printed in Table II column 5. It can be easily done and the interpolating polynomial
is presented by
\begin{equation}
\label{P4-Li-s}
 D_{rQED} = (-1.696 + 2.103\, Z - 1.272\, Z^2 + 0.5215\, Z^3 - 0.1536\, Z^4)\times 10^{-4} \ .
\end{equation}
It reproduces $D_{rQED}$ estimates in all points in $Z$ except for $Z=8, 16, 20$ where it differs in one unit
in the 3rd figure. It implies that the results for approximate method used in \cite{MTS:2018} are modeled by
(\ref{P4-Li-s}) with high accuracy. As the second goal we want to approximate the known reliable results for
$D_{rQED}$ for $Z=3$ \cite{Drake:2003,Pachucki:2008} and $Z \geq 10$ \cite{MTS:2018}. Surprisingly, the polynomial,
\begin{equation}
\label{P4-Li-t}
 D_{rQED}^{\rm Li-like} = (-37.22 + 26.33\, Z - 5.925\, Z^2 + 0.8735\, Z^3 - 0.1629\, Z^4)\times 10^{-4}\ ,
\end{equation}
reproduces $D_{rQED}$ results in 3 s.d. for $Z=3$ and $Z \in [10,20]$, see Table II, column 6.
Deviations which occur for $Z \in [4, 9]$ measure an inaccuracy of the method used in \cite{MTS:2018}.

\section{Variational energies vs exact ones}

\subsection{\it Two-electron case}

\begin{figure}[tb]
  \begin{center}
   \includegraphics*[width=3in,angle=-90]{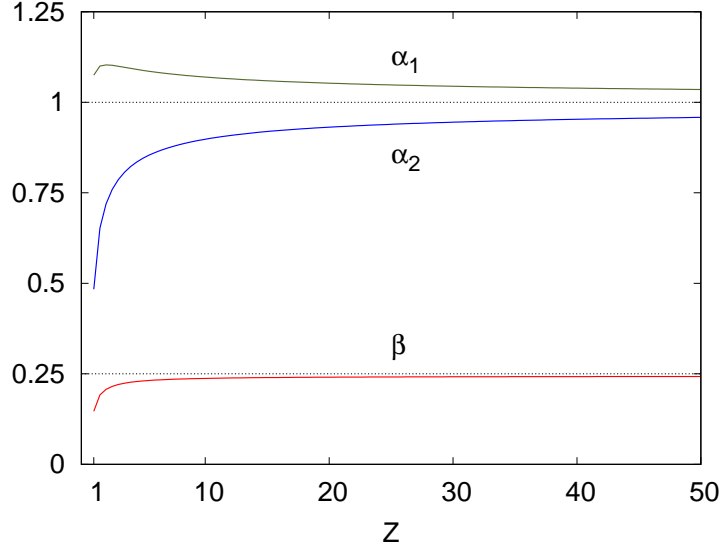}
    \caption{
    \label{Figure2}
    Variational Parameters {\it vs} the nuclear charge $Z$ for the two-electron system using
    the trial function
    \hbox{$\psi=(e^{-\al_1 Z r_1 - \al_2 Z r_2} + e^{ - \al_2 Z r_1 - \al_1 Z r_2})\,e^{\beta r_{12}}$},
    see (\ref{HeTrial}), assuming $\al_1 \geq \al_2$.
    }
  \end{center}
\end{figure}

In 1929 E Hylleraas in his celebrated paper \cite{Hylleraas}  proposed to use for Helium type system
$(2e; Z)$ the exponentially correlated trial function in the form of symmetrized product of three
(modified by screening) Coulomb Orbitals (cf. formula (13) in Ref.\cite{Hylleraas}),
\begin{equation}
\label{HeTrial}
  \Psi_0\ =\ (1 + P_{12})\, e^{\scriptstyle - \al_1 Z r_1 - \al_2 Z r_2 + \beta r_{12}}\ ,
\end{equation}
with three variational parameters $\al_{1,2} , \beta$, here $P_{12}$ is permutation operator $1 \lrar 2$.
Many years after, Calais-L\"owdin \cite{CL} demonstrated that all integrals involved to the variational
calculations are intrinsically 3-dimensional in $r_1, r_2, r_{12}$ variables (for discussion see \cite{twe})
and they can be evaluated analytically. Eventually, the variational energy is a
certain rational function of parameters $\al_{1,2} , \beta$.   Hence, the procedure of minimization
of the variational energy is essentially algebraic and can be easily performed. On Fig.\ref{Figure2}
the optimal parameters {\it vs} the nuclear charge $Z$ are presented - they are smooth, slow-changing
functions. At $Z$ tends to infinity the $\al$-parameters approach slowly to one, $\al_{1,2} \rar 1$,
and $\beta \rar 1/2$, while asymptotically, at $Z \rar \infty$ the function (\ref{HeTrial})
(in appropriate variables) becomes the product of two ground state Coulomb orbitals.
Making concrete calculations for different
values of $Z$ one can see that the variational energy coincides systematically with exact NRQED energies in 4-3 s.d.
while at $Z=2-12$, in fact, it differs in the 3rd d.d.! Thus, the simple trial function (\ref{HeTrial}) describes
the energy in the domain of applicability of NRQED in static approximation. Overall quality
of the trial function (\ref{HeTrial}) can be "measured" by how accurately it reproduces
the electron-nucleus cusp parameter $Z$ (the residue in Coulomb singularity at $r_1=0$ or $r_2=0$) by
$\frac{(\al_1+\al_2)Z}{2}$, see Fig.\ref{Figure2}. If at small $Z$ the difference is of order 10$\%$,
then it reduces to 0.01$\%$ at $Z=12$ and tends to zero at large $Z$. For reasons unclear to the authors
the electron-electron cusp at $r_{12}=0$ is not well-reproduced, it differs in about 50$\%$. It means that the variational trial function
does not behave correctly in vicinity of $r_{12}=0$, which however does not influence the quality of variational energy.
This question will be studied elsewhere.

\subsubsection*{Two-electron case: effective potential}

Taking trial function $\Psi_0$ (\ref{HeTrial}) with optimal parameters one can calculate a potential
for which this function is the exact ground state function
\begin{equation}
\label{Veff2e}
 V^{(2e)}_{eff}\ =\ \frac{\De \Psi_0}{\Psi_0}\ ,
\end{equation}
which we will call the {\it effective} potential for two-electron problem. It can be easily checked that this potential
reproduces Coulomb
singularities at $r_1=0, r_2=0,
r_{12}=0$\,.
One can define the effective theory with Hamiltonian,
\begin{equation}
\label{Heff2e}
 {\cal H}_{eff}\ =\ -\frac{1}{2} \sum_{i=1}^2 \De_i\ +\ V^{(2e)}_{eff}\ ,
\end{equation}
for which the ground state energy coincides with NRQED energy in its domain of applicability.

Taking $\Psi_0$ as the zero approximation in Non-Linearization Procedure with
$V_0=V^{(2e)}_{eff}$ as unperturbed potential, one can develop the convergent perturbation theory w.r.t.
difference between the original potential
$V$ (\ref{H}) and $V_0$ (\ref{Veff2e}), see for review \cite{Turbiner:1984}. The sum of the first
two terms $(E_0+E_1)$ coincides with the variational energy with trial function $\Psi_0$. It is evident
that the next correction $E_2$ is the first
quantum correction to NRQED; in general, it changes the 3rd (and higher) d.d. in the variational energy.
This procedure allows us to calculate
quantum corrections
to NRQED with static nuclei. These corrections, of course, can be calculated indirectly using variational
method by taking more complicated trial
functions than (\ref{HeTrial}), in particular, their linear superpositions, see e.g. \cite{Korobov:2000}.

\begin{figure}[tb]
 \begin{center}
   \includegraphics*[width=3in,angle=-90]{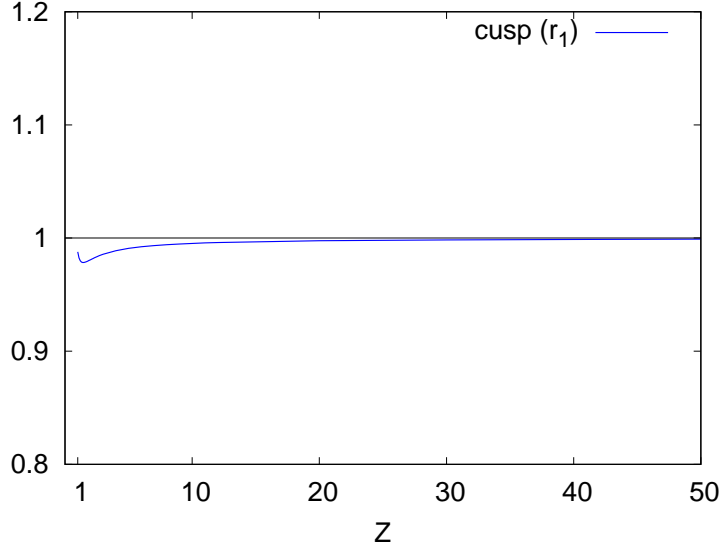}
    \caption{
    \label{Figure3}
    (Normalized) variational electron-nucleus cusp parameters (the residues at Coulomb singularities
    in (\ref{H})): $\mbox{cusp}(r_1)=\mbox{cusp}(r_2)$
    {\it vs} $Z$ for the two-electron system using the trial function  (\ref{HeTrial}),
    the exact cusp value is 1.
    }
  \end{center}
\end{figure}

\subsection{\it Three-electron case.}

For Lithium-type system $(3e; Z)$ let us take a variational trial function
(for the total spin 1/2) in the form
\begin{equation}
\label{GStrialfunct}
 \psi(\vec{r}_1,\vec{r}_2,\vec{r}_3; \chi) = {\cal A} \left[ \,
   \phi(\vec{r}_1,\vec{r}_2,\vec{r}_3)\chi \, \right]\,,
\end{equation}
where $\chi$ is the spin eigenfunction, ${\cal A}$ is
the three-particle antisymmetrizer
\begin{equation}\label{Asym}
 {\cal A} = 1 - P_{12} - P_{13} - P_{23} + P_{231}  + P_{312}\,,
\end{equation}
and  $\phi(\vec{r}_1,\vec{r}_2,\vec{r}_3)$ is the explicitly
correlated orbital  function
\begin{equation}
\label{Phi3funct}
\phi(\vec{r}_1,\vec{r}_2,\vec{r}_3; \al_i, \al_{ij})  \ =\
 e^{\scriptstyle -\al_1 Z r_1-\al_2 Z r_2-\al_3 Z r_3} \,
e^{\scriptstyle \al_{12} r_{12} + \al_{13} r_{13} + \al_{23} r_{23}} \ ,
\end{equation}
see e.g. \cite{TGH2009}, where $\al_i$ and $\al_{ij}$ are non-linear variational
parameters. Here, $P_{ij}$ represents the permutation $i \leftrightarrow j$, and
$P_{ijk}$ stands for the permutation of $(123)$ into $(ijk)$.
In total, (\ref{GStrialfunct}) contains
six all-non-linear variational parameters.  The function (\ref{GStrialfunct})
is a properly anti-symmetrized product of $(1s)$ (modified by screening)
Coulomb orbitals and the exponential correlation factors $\sim\exp{(\al_{ij}\, r_{ij})}$.

There are two linearly independent spin $1/2$ functions of mixed symmetry:
\begin{equation}
\label{chi1}
 \chi_1\ =\  \frac{1}{\sqrt{2}} [ {\boldsymbol\alpha}(1)
   {\boldsymbol\beta}(2)    - {\boldsymbol\beta}(1)
   {\boldsymbol\alpha}(2) ]{\boldsymbol\alpha}(3)
\end{equation}
and
\begin{equation}
\label{chi2}
 \chi_2\ =\  \frac{1}{\sqrt{6}} [  2{\boldsymbol\alpha}(1)
   {\boldsymbol\alpha}(2)  {\boldsymbol\beta}(3)   -
   {\boldsymbol\beta}(1)  {\boldsymbol\alpha}(2)
   {\boldsymbol\alpha}(3)  - {\boldsymbol\alpha}(1)
   {\boldsymbol\beta}(2) {\boldsymbol\alpha}(3) ] \,,
\end{equation}
where ${\boldsymbol\alpha}(i)$, ${\boldsymbol\beta}(i)$ are spin up,
spin down eigenfunctions of $i$-th electron, respectively. For simplicity, the spin function
$\chi$ in (\ref{GStrialfunct}) is chosen as
\begin{equation}
\label{chi}
   \chi\ =\ \chi_1 + c\, \chi_2 \ ,
\end{equation}
(see \cite{TGH2009}), where $c$ is a variational parameter. It implies that the coordinate (orbital) functions
in front of $\chi_{1,2}$ are the same.
Eventually, the trial function is a linear superposition of twelve terms, it contains 7 free parameters.

The variational energy is given by the ratio of two nine-dimensional integrals.
In relative space coordinates
$(r_1, r_2, r_3, r_{12}, r_{13}, r_{23}, \Om(\tha_1, \tha_2, \tha_3))$
the integration over three angles $\Om$ describing overall orientation and rotation of the system
are easily performed analytically. We end up with six-dimensional integrals over the relative distances
$(r_1, r_2, r_3, r_{12}, r_{13}, r_{23})$ (for the general discussion see \cite{twe}).
It was shown a long ago by Fromm and Hill \cite{Fromm-Hill} that these integrals
can be reduced to one-dimensional ones (!) but with integrands involving dilogarithm functions.
The analytic properties of the resulting expressions for integrands are found to
be unreasonably complicated (see e.g. \cite{Harris}) for numerical evaluation.
For that reason, the method we used is direct numerical evaluation
of the original six-dimensional integrals.

\begin{figure}[tb]
  \begin{center}
   \includegraphics*[width=2.in,angle=-90]{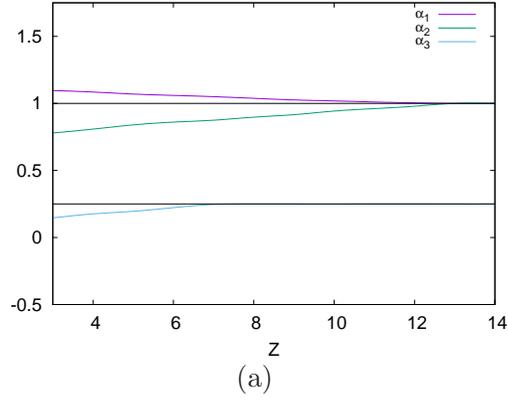} \\ (a) \\
   \includegraphics*[width=2.in,angle=-90]{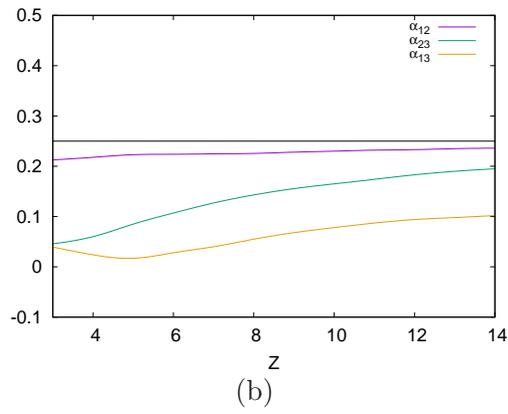} \\ (b) \\
   \includegraphics*[width=2.in,angle=-90]{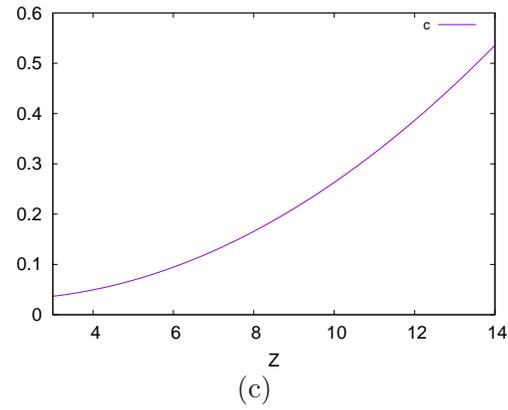} \\ (c)
    \caption{
    \label{Figurexa}
    Variational Parameters {\it vs} the nuclear charge $Z$ for the three-electron system using the trial function
(\ref{GStrialfunct})-(\ref{Phi3funct}): $\al_{1,2,3}$ (a), $\al_{12,23,13}$ (b) and $c$ (c), see Eq. (\ref{chi})
    }
  \end{center}
\end{figure}

The results of variational calculations are shown in Table \ref{Interpolation3e}, last column
for $Z=3 - 14$. Variational parameters {\it vs} $Z$ are presented in Fig. \ref{Figurexa},
they are smooth, slow-changing functions. Parameter $c$ being small at $Z=3,4,\ldots$ grows with $Z$
reaching sufficiently large value $~0.536$ at $Z=14$. It indicates the importance of the contribution
emerging from the second spin function $\chi_2$
(\ref{chi2}) for large $Z$.
It might be considered as the indication that the condition (\ref{chi}) should be relaxed
and the orbital functions should be different,
\begin{equation}
\label{chi-complete}
   \phi\,\chi\ =\ \phi(\vec{r}_1,\vec{r}_2,\vec{r}_3; \al_i^{(1)}, \al_{ij}^{(1)})\,\chi_1 +
   c\,\phi(\vec{r}_1,\vec{r}_2,\vec{r}_3; \al_i^{(2)}, \al_{ij}^{(2)})\, \chi_2 \ ,
\end{equation}
see (\ref{GStrialfunct}), where $\phi$'s are given by (\ref{Phi3funct}).
Now this trial function depends on 13 free parameters. Immediate calculation shows a significant improvement
in variational energy even for $Z=3$: -7.471 a.u. {\it vs} -7.455 a.u., see Table II \cite{delValle-Nader}.

It is easy to check that variational energies reproduce not less than 99.9$\%$ of the exact non-relativistic
energies, hence, the NRQED energies in
static approximation. Moreover, relaxing condition (\ref{chi}), thus, assuming that both orbital functions
are different being the type
(\ref{Phi3funct}), should increase accuracy. Note that the overall quality of the trial function
(\ref{GStrialfunct})-(\ref{chi}) is reflected in accurate reproduction of electron-nuclear cusp parameter
2.953 \cite{TGH2009} at $Z=3$ while the exact one is equal to 3, thus, it deviates in about 2$\%$.
Using the function (\ref{chi-complete}) the value of electron-nuclear cusp parameter gets even better
2.991 \cite{delValle-Nader}.

Similar to two-electron case for reasons unclear to the authors the electron-electron cusp at $r_{ij}=0$
is not well-reproduced being smaller in about 50 $\%$. It means that the variational trial function
does not behave correctly in vicinity of $r_{ij}=0$, which it does not influence the quality
of variational energy. This question will be studied elsewhere.

\subsubsection*{Three-electron case: effective potential}

Taking trial function $\Psi_0$ (\ref{GStrialfunct}) with optimal parameters one can calculate a potential
for which this function is the exact ground state function
\begin{equation}
\label{Veff3e}
 V^{(3e)}_{eff}\ =\ \frac{\De \Psi_0}{\Psi_0}\ ,
\end{equation}
cf. (\ref{Veff2e}), which we will call the {\it effective} potential for three-electron problem.
It can be easily checked that this potential
reproduces Coulomb singularities
at $r_1=0, r_2=0, r_3=0, r_{12}=0, r_{23}=0, r_{13}=0$\,. One can define the effective theory with Hamiltonian,
\begin{equation}
\label{Heff3e}
 {\cal H}_{eff}\ =\ -\frac{1}{2} \sum_{i=1}^3 \De_i\ +\ V^{(3e)}_{eff}\ ,
\end{equation}
for which the ground state energy coincides with NRQED energy in its domain of applicability.


\section*{\large Conclusions}

Concluding we state that a straightforward interpolation between small and
large $Z$ in a suitable variable $\la$ (\ref{lambdadef}) based on a meromorphic
function $\mbox{gPade}(9/5)_{3,4}\,(\la(Z))$ (10) leads to accurate description of
12-13-14 s.d. of the non-relativistic ground state energy of both the Helium-like and
Lithium-like ions in static approximation, thus, for $1s^2\ {}^1 S$ and $1s^2\, 2s \ {}^2S$
states, respectively, for $Z \leq 20$. Even at critical charge $Z_c^{(2)}$ for Helium-like
system it provides 10 s.d. in ground state energy correctly. In general, such an outstanding
accuracy provided by (10) leads to a hint that the Puiseux expansion (5) at $Z=Z_B$ has
finite radius of convergence while the expansion at $Z=Z_c$ is a Taylor expansion
with finite radius of convergence as well. It also indicates that the radius of convergence
of $1/Z$-expansion is defined by $Z_B$.

It is worth noting that the quality of approximation increases substantially with the growth
of $N$ in generalized two-point Pade approximation (9): at $N=0$ it gives accuracy of 3-4 s.d.
in the ground state energy for all $Z$ studied, while at $N=4$ the accuracy increases to 7-8 s.d.
and, eventually, at $N=5$ it gives the above-mentioned accuracy 12-13-14 s.d.
It seems natural to assume that the similar interpolation has to provide
high accuracy for the energies of excited states of above systems and
even for other many-electron atomic systems. It will be presented elsewhere
\cite{TLOVN:2017}.

It has to be emphasized that at present for Helium atom the experimental accuracy
is of the order $10^{-11}$ in relative units \cite{Nature:2014}
(for discussion see \cite{Pachucki:2017}). Theoretically, the contributions of the
order of $\alpha^8$ $(\alpha^7$ for excited states) are unknown. It seems important to know
to what significant digit in ground state energy this correction will give contribution.
It might be that this correction as well as non-QED contributions e.g. hadronic loops may
influence 9-10th significant digit in energy. Let us emphasize that our interpolation provides
non-relativistic energies well beyond the present accuracy of both experimental and theoretical data.

Note that similar interpolation works very well for simple diatomic molecules
H$^+_2$, H$_2$, He$_2^+$ and ${\rm HeH}$ in Born-Oppenheimer approximation
matching perturbation theory at small internuclear distances and multipole
expansion with instanton-type, exponentially-small contributions at
large distances (as for the first three systems).
It provides 4-5-6 s.d. at potential curves depending on internuclear distances
and eventually not less than 5-6 s.d. for spectra of existing rovibrational
states \cite{OT:2017}.

Making detailed analysis of finite mass corrections, QED and relativistic effects for
the ground state energy of 2- and 3-electron ions for $Z \leq 20$ we
localized the domain of applicability of NRQED in static
approximation. This domain is limited by 4-3 s.d. in the ground state energy
for both systems. Surprisingly, this domain is described accurately by a 4th
degree polynomial (without linear term) in variable ${\la}=\sqrt{Z-{Z_B}}$,
where $Z_B$ is the 2nd critical charge \cite{TLO:2016}.
This domain can also be fitted by the Majorana formula - the 2nd degree
polynomial in $Z$ (\ref{majorana}) with two free parameters, while $e_2$ - the
coefficient in front of $Z^2$ term - is kept fixed and equal to the sum of the
(ground state) energies of 2(3)-Hydrogen atoms  - with similar accuracies of 4-3 s.d.!
Remarkably, the first 3 s.d. in the leading approximation of the sum of relativistic
and QED corrections are systematically described by 4th degree polynomial in Z for
$Z \leq 50$ for Helium-like and for $Z \leq 20$ for Lithium-like systems.

Note that the Majorana formula (\ref{majorana}) (as well as generalized two-point
Pade approximations) allows us to calculate the (first) critical charge in
reasonably accurate way. Adding to the Majorana formula the
$\la^3$ term slightly improves the quality of approximation at small
integer $Z$ but allows to describe correctly the energy in vicinity of the
first (second) critical charge $(Z_c) Z_B$. Striking fact is all three curves
shown on Fig.1 are, in fact, parabolas (up to width of drawing line)!

It seems interesting to check applicability of Majorana formula for NRQED with
finite mass nuclei. In order to do it we calculated for the first time in full
geometry the non-relativistic ground state energy of two-electron atomic system
for $Z \in [11, 20]$ in Lagrange mesh method with accuracy not less than 10 d.d.
It complements the results by Nakashima-Nakatsuji \cite{Nakashima:2008} for
$Z\in [1,10]$. These results are displayed in the 3rd column of Table I. As for 3-electron systems
one-two leading finite-mass corrections are included
into the ground state energy, see Table II, 3rd column. The sum of relativistic and QED
corrections in leading approximation remains almost unchanged. Domain of applicability of
NRQED is again limited to 3-4 s.d. In both cases of two and three electron sequences
the Majorana formula - the second degree polynomial in $Z$ - continues to describe
domain of applicability of NRQED with slightly changed coefficients, see Table V and VI.

Interestingly, making a generalization of the Slater determinant method by
including inter-electronic correlations in exponential form, thus,
taking a trial function in the form of (anti)-symmetrized product of three (six)
modified-by-screening Coulomb orbitals for two-(three-) electron system
(they can be called {\it generalized Hylleraas} functions), respectively,
allows us to reproduce $\sim 99.9\%$ of the ground state energy from small
$Z$ up to $Z=20$. Of course, it requires a careful minimization with respect
to screening (non-linear) parameters.
In other words, the obtained variational energy, in fact, coincides with
exact energy in domain of applicability of NRQED with (in)finitely-heavy nuclei.
This observation hints that such a generalized Slater determinant method might
be successful for other atomic and molecular systems. It will be checked elsewhere.

\begin{acknowledgments}
A.V.T. is grateful to Physics Department of Stony Brook University and Simons Center
(Stony Brook, USA) where some parts of the present work were carried out.
J.C.L.V. thanks PASPA grant (UNAM, Mexico) and the Centre de Recherches Math\'ematiques,
Universit\'e de Montr\'eal, Canada for the kind hospitality while on sabbatical leave
during which this work was initiated. H.O.P. wants to express a deep gratitude to ICN-UNAM (Mexico),
where the essential part of the present work was done during his numerous visits.
A.V.T. thanks M.I.~Eides, V.M.~Shabaev and V.A.~Yerokhin for useful remarks.
The authors thank G.W.F.~Drake for important discussions and clarifications.
Most of calculations associated with generalized Slater determinants - generalized
Hylleraas functions were carried out on 120-processor cluster KAREN
(ICN-UNAM, Mexico). In the last stage the research was supported partially by
CONACyT grant A1-S-17364 and DGAPA grant IN108815 (Mexico).
The authors express deep gratitude to two anonymous referees for careful reading
of the manuscript, constructive suggestions and critical remarks -- all that helped
to improve the presentation.
\end{acknowledgments}

\end{document}